\documentclass[aps,twocolumn,showpacs]{revtex4}
\usepackage{graphics}
\usepackage{graphicx}
\usepackage{amsmath}
\usepackage{amsfonts}
\usepackage{amssymb}
\usepackage{dsfont}
\def\beqn{\begin{eqnarray}}
\def\eeqn{\end{eqnarray}}
\def\aap{a^\dagger}
\def\aam{a}
\def\hw{\hbar \omega}
\def\hw4{ \frac {\hbar \omega}{4}}

\def\uni{{\bf i}}
\def\re{{\rm {e}}}
\def\rd{{\rm {d}}}

\def\al{\alpha}
\def\be{\beta}
\def\sg{\sigma}
\def\Om{\Omega}
\def\om{\omega}
\def\b0{b_0}
\def\wid{\widetilde}
\def\ovl{\overline}
\def\hud{\hat{u}^\dagger}
\def\hu{\hat{u}}
\def\hvd{\hat{v}^\dagger}
\def\hv{\hat{v}}
\def\inte{\int_{-\infty}^\infty}
\begin{document}

\title{Swanson Hamiltonian: non-PT-symmetry phase.}

\vskip2cm
\author{Fern\'andez V. $^{a)}$}
\author{Ram\'\i rez R. $^{b)}$}
\author{Reboiro M.     $^{c)}$}

\affiliation{{\small\it $^{a)}$} Department of Mathematics, University of La Plata, Argentina}
\affiliation{{\small\it $^{b)}$} IAM, CONICET-CeMaLP, University of La Plata, Argentina}
\affiliation{{\small\it $^{c)}$}IFLP, CONICET-Department of Physics, University of La Plata, Argentina}

\date{\today}

\begin{abstract}
In this work, we study the non-hermitian Swanson hamiltonian, particularly the non-PT symmetry phase. 
We use the formalism of Gel'fand triplet to construct the generalized eigenfunctions and the corresponding spectrum. Depending on the region of the parameter model space, we show that the Swanson hamiltonian represents different physical systems, i.e. parabolic barrier, negative mass oscillators. We also discussed the presence of Exceptional Points of infinite order. 
\end{abstract}

\pacs{21.60.-n; 21.60.Fw; 02.20.Uw.}

\maketitle

keywords: PT-symmetry Swanson Hamiltonian, Gel'fand triplet, Exceptional Points.

\section{Introduction.}

The study of non-hermitian Parity-Time Reversal(PT) symmetry hamiltonians was proposed in the pioneering work of Bender and Boettcher \cite{bender0}. A parametric family of hamiltonians can be obtained by varying the different variables of the corresponding systems. The essential feature of a PT-symmetric Hamiltonian is the existence of certain values of the parameters at which the spectrum is real and the hamiltonian is similar to a hermitian one \cite{bender1,bender2,bender3,bender4,bender5,bender6}. For other values of the parameters of the model, the spectrum contains complex-conjugate pairs of eigenvalues and the corresponding eigenfunctions are non longer PT-symmetric. In the boundary of both regions of the model space, two or more eigenvalues and their corresponding eigenstates can be coalescent, these set of parameters are called Exceptional Points \cite{ep1,ep2,ep3,ep4,ep5,ep6,ep7,ep8,ep9,ep10,znojil-pe1,znojil-pe2,znojil-pe3}. The time evolution of a given initial state under the action of these hamiltonians strongly depends on the characteristics of the spectrum, particular at EPs \cite{garmon1,garmon2,garmon3,garmon4,hatano1,hatano2,hatano3,
hatano4,nos1,nos2}, where the exponential law is generally not valid.  

The Swanson model has been introduced in \cite{swanson} as an example of a hamiltonian that obeys PT-symmetry. 
It admits real eigenvalues for a well-defined region of the parameter model space. The similarity between the Swanson hamiltonian and the harmonic oscillator as well as the dynamic of observables in the PT-symmetry region have been extensively analyzed \cite{ahmed1,ahmed2,geyer,quense,sinha-sw1,sinha-sw2,sinha-sw3,sw-1,sw-2,fring-sw,sw-close1,mostaf-sw,mostaf-general,znojil-sw-inner,fring}. Among the extensions of the Swanson model we can include super-symmetry realizations \cite{sinha-sw-susy,sw-susy}. Also, the Swanson hamiltonian has been studied as a particular case of different non-hermitian generalizations of anharmonic oscillators systems \cite{znojil-osc1,znojil-osc-exp,znojil-anarm-osc,znojil-osc2}. Another approach to the Swanson model comes from investigating different q-deformation boson algebras \cite{fring-sw3,fring-sw1,fring-sw2,nos3}. As an example, in \cite{fring-sw3}
different representations of deformed canonical variables \cite{fring-sw3} are presented. In this work, the effect that produces the modification of the generalized canonical variables commutator, over the region of PT-symmetry and exceptional points, is analyzed. In \cite{fring-sw1}, the authors studied the particular case of the Swanson hamiltonian using the generalization of the Milne quantization. In \cite{fring-sw2} the Swanson hamiltonian is discussed in the framework the formalism of generalized pseudo bosons. The authors of \cite{fring-sw2}, by employing generalized Bogoliubov transformation, present a mapping of the Swanson model to a standard bosons hamiltonian. In \cite{nos3} the Swanson model is obtained as the quadratic limit of a deformed general hamiltonian constructed from a non-standard oscillator algebra. 

However, to our knowledge, much less has been investigated in the region of PT-broken symmetry. In this line, the authors of \cite{sinha-continuum} have proposed different generalizations of the Swanson model. They have described the continuum spectrum of the different generalizations by analyzing the corresponding similar hermitian hamiltonians. More recently, the authors of \cite{fring-sw4} have studied the solutions of the time-dependent Swanson hamiltonian. By applying the formalism presented in \cite{fring-sw6}, their work includes the construction of a time-dependent metric to compute the time evolution of the observables of the system. A new proposal has been presented in \cite{fring-sw5}, where the authors construct time-dependent metrics by point transformations. In this work, the construction of non-Hermitian invariants for the Swanson model and the implementation of Dyson maps is analyzed. From another point of view, the authors of \cite{mihail-darboux} have made use of the Darboux transformation to provide solutions of the Swanson model for a particular set of parameters.

In this work, we study the non-hermitian Swanson hamiltonian, particularly the non-PT symmetry phase.
 The work is organized as follows. In Section \ref{formal} we analyse the different model space regions. In Subsection \ref{gelfand} we discuss the use of the formalism of Rigged Hilbert Space to construct the generalized eigenfunctions and the corresponding spectrum of the model. In Subsection \ref{regions} we present our results for each region of the model parameter space. In Section \ref{evolution}  we formalize the calculation of mean values of observables and their time evolution. Conclusions are drawn in Section \ref{conclu}.

\section{Formalism.}
\label{formal}

Let us start with the hamiltonian of the squeezed harmonic oscillator \cite{sho,sho2,sho3}.  It is given by
\begin{equation} 
H= \hbar \omega~ \left( \aap \aam+ \frac 12 \right) + 
\hbar \alpha~ ({\aam}^2+ ~ {\aap}^2).
\label{hsq}
\end{equation}
As it is well-known \cite{sho,sho2,sho3}, its relevance is related to the study of the Heisenberg Uncertainty Relations for the momentum and position operators. The eigenvalues and eigenfunctions of the hamiltonian of Eq.(\ref{hsq}),  for $|\alpha|<|\omega|/2$, can be obtained analytically, and the lowest eigenstate is a squeezed state, that is a state which minimizes the variance of the momentum operator $\hat p$ by increasing the variance of the coordinate operator $\hat x$, or vice versa. 

Let us consider the Hamiltonian proposed in \cite{swanson}, which is a non-hermitian generalization of the hamiltonian of Eq.(\ref{hsq}). It reads

\begin{equation}
H(\omega,\alpha,\beta)= \hbar \omega~ \left( \aap \aam+ \frac 12 \right) + 
\hbar \alpha~ {\aam}^2+  \hbar \beta ~ {\aap}^2 .
\label{hswsq}
\end{equation}

We can write $\aam$ and $\aap$ in terms of the coordinate operator, ${\hat x}$, and of the momentum operator, ${\hat p}:$

\beqn
\aam & = & \frac{1}{\sqrt{2}} \left ( \frac {\hat x} {b_0} + {\bf i} \frac {b_0} {\hbar} \hat{p}\right), \nonumber \\
\aap & = & \frac{1}{\sqrt{2}} \left ( \frac {\hat x} {b_0} - {\bf i} \frac {b_0} {\hbar} \hat{p}\right), \nonumber \\
\eeqn
with $b_0$ the characteristic length of the system. The hamiltonian of (\ref{hswsq}) reads

\beqn
H (\omega,\alpha,\beta) & = & 
\frac 1 2 \hbar (\omega + \alpha +\beta ) \left( \frac {\hat x} {b_0}\right)^2 \nonumber \\
& &+ \frac 1 2 \hbar (\omega - \alpha -\beta ) \left( \frac {b_0~ \hat{p}} \hbar \right)^2
\nonumber \\
& &+  \hbar \frac{(\alpha-\beta)}{2} \left( 2~\hat{x} \frac{ {\bf i}}{\hbar} \hat{p}+1 \right).
\label{hxp}
\eeqn

It is straightforward to verify that the hamiltonian of Eq.(\ref{hxp}) obeys PT-symmetry. 

In the other hand, the hermitian conjugate operator, $H_c$, is given by
$H_c(\omega,\alpha,\beta)=H(\omega,\beta,\alpha)$.

Let us introduce, for $\omega \neq \alpha + \beta$, a new set of complementary operators $\hat {P}$ and $\hat X$ , namely: 

\beqn
\hat{P} & = &  \left (
{\hat p} +
\uni \hbar 
\frac {\alpha-\beta}{(\omega-\alpha-\beta)b_0^2}  {\hat x} \right ),\nonumber \\
\hat{X} & = &  {\hat x},
\label{PX}
\eeqn
$\hat {P}$ and $\hat X$ obey the usual commutation relation 
 $[\hat{X},\hat{P}]=\nolinebreak \uni\hbar.$

In terms of ${\hat{P}}$ and $\hat{X}$ the Hamiltonian of Eq.(\ref{hxp}) can be written as

\beqn
H= \frac 1 {2 m}~\hat{P}^2 \phi(X) + \frac k 2 ~ {\hat X}^2 \phi(X),
\label{newosc}
\eeqn
with $k= m ~ \Omega^2$ and 

\beqn
\Omega=\Omega(\omega,\alpha,\beta)& = & \sqrt{\omega^2-4 \alpha \beta}=|\Omega| {\rm {e}}^{{\bf i} \phi},
\label{energy}
\eeqn

\beqn
m=m(\omega,\alpha,\beta,b_0)& = &\frac{ \hbar}{(\omega-\alpha-\beta) b_0^2}.
\label{masseff}
\eeqn

The spectrum and the eigenfunctions of $H$ depend on the sign of the functions $m(\omega,\alpha,\beta,b_0)$, $\Omega(\omega,\alpha,\beta)^2$. 

In Figures 1 and 2 we show the behaviour of $\Om^2$ and $m$,  as a function of $\al/\om$ and $\be/\om$, respectively. It can be observed that $\Om^2$ is a continuous function, while $m$ has a discontinuity at the plane $\al/\om+\be/\om-1=0$. 

\begin{figure}
\includegraphics[width=8cm]{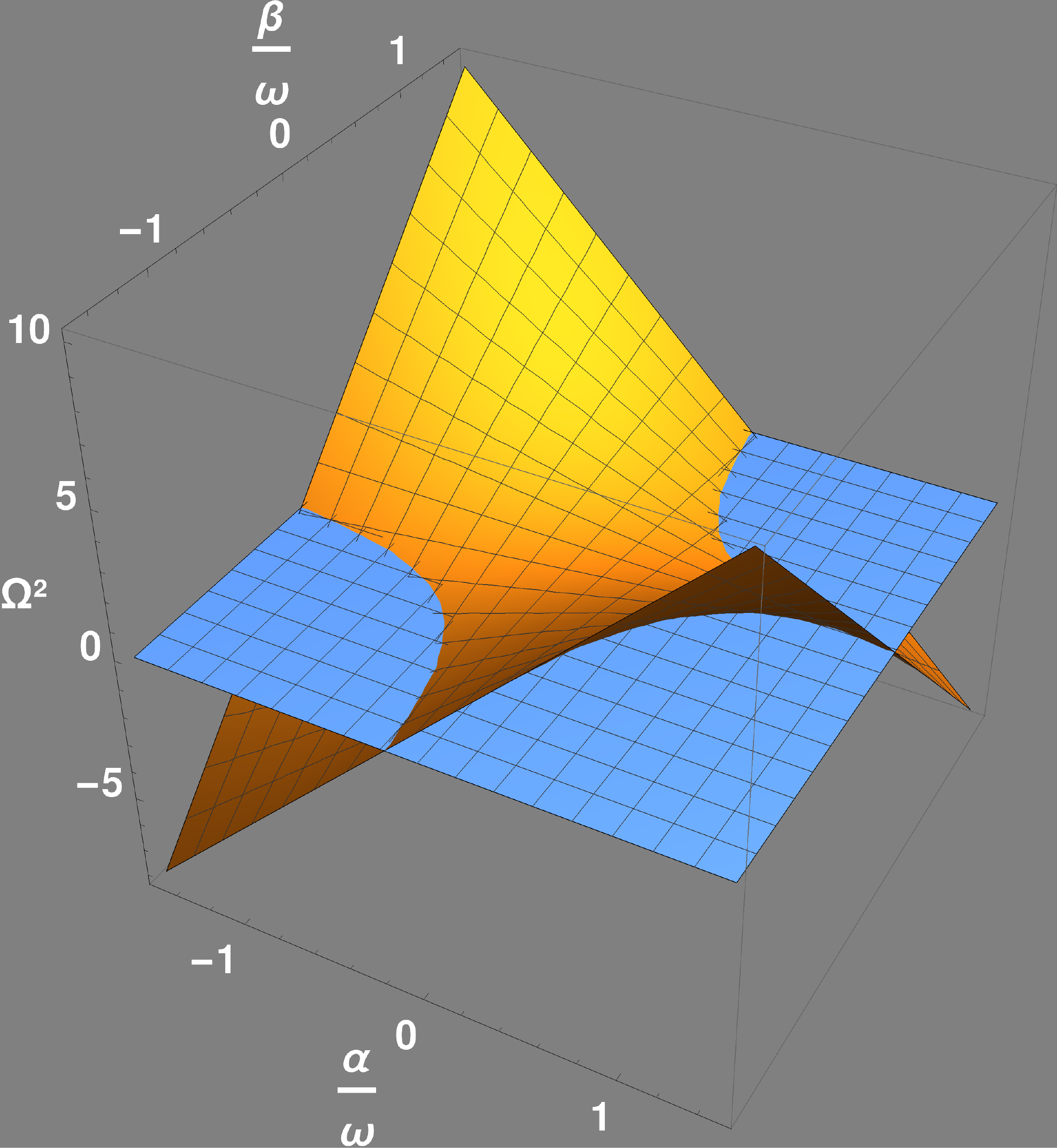}
\caption {$\Omega^2(\omega,\alpha,\beta,b_0)$ of Eq.(\ref{energy}). }\label{fig:fig1}
\end{figure}

\begin{figure}
\includegraphics[width=8cm]{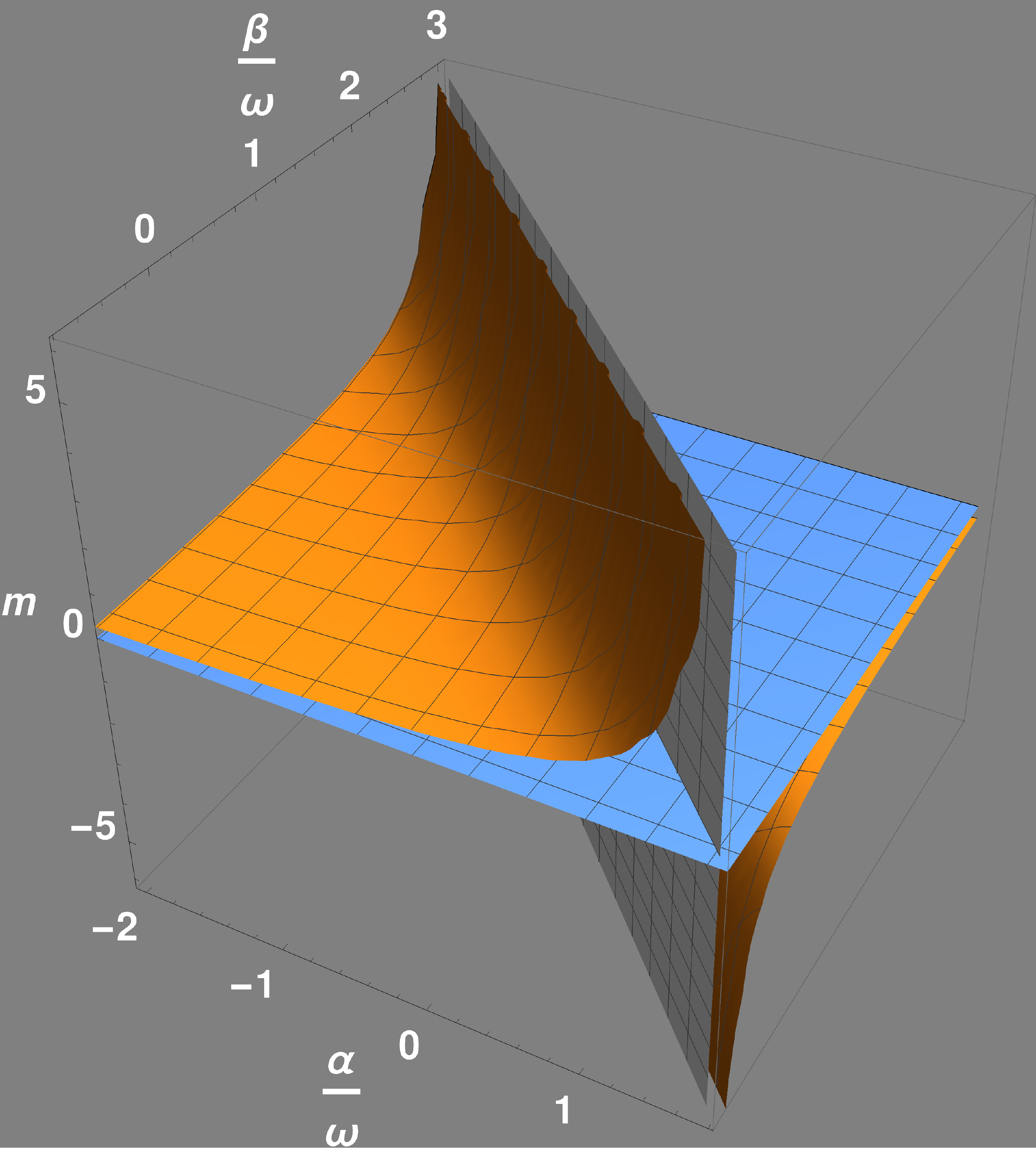}
\caption {$m(\omega,\alpha,\beta,b_0)$ of Eq.(\ref{masseff}). }\label{fig:fig2}
\end{figure}

There are four possible regions in the parameter model space. If $m>0$ and $\Omega^2>0$ the hamiltonian is similar to the usual harmonic oscillator, Region I.  The case $m>0$ and $\Omega^2<0$ corresponds to a parabolic barrier, Region II\cite{chu1,chu2,maru,bermudez}. While the case $m<0$ and $\Omega^2>0$ can be interpreted as a harmonic oscillator with negative mass, Region III \cite{neg-mass-1,neg-mass-2,neg-mass-3,neg-mass-4,neg-mass-5}. Finally, the case $m<0$ and $\Omega^2<0$ can be interpreted as a parabolic barrier for a system with negative mass, Region IV. The case $k(\omega,\alpha,\beta,b_0)=0$ corresponds to a free particle. The different regions are displayed in Figure 3 in terms of the adimensional coupling constants $\alpha/\omega$ and $\beta/\omega$.

\begin{figure}
\includegraphics[width=8cm]{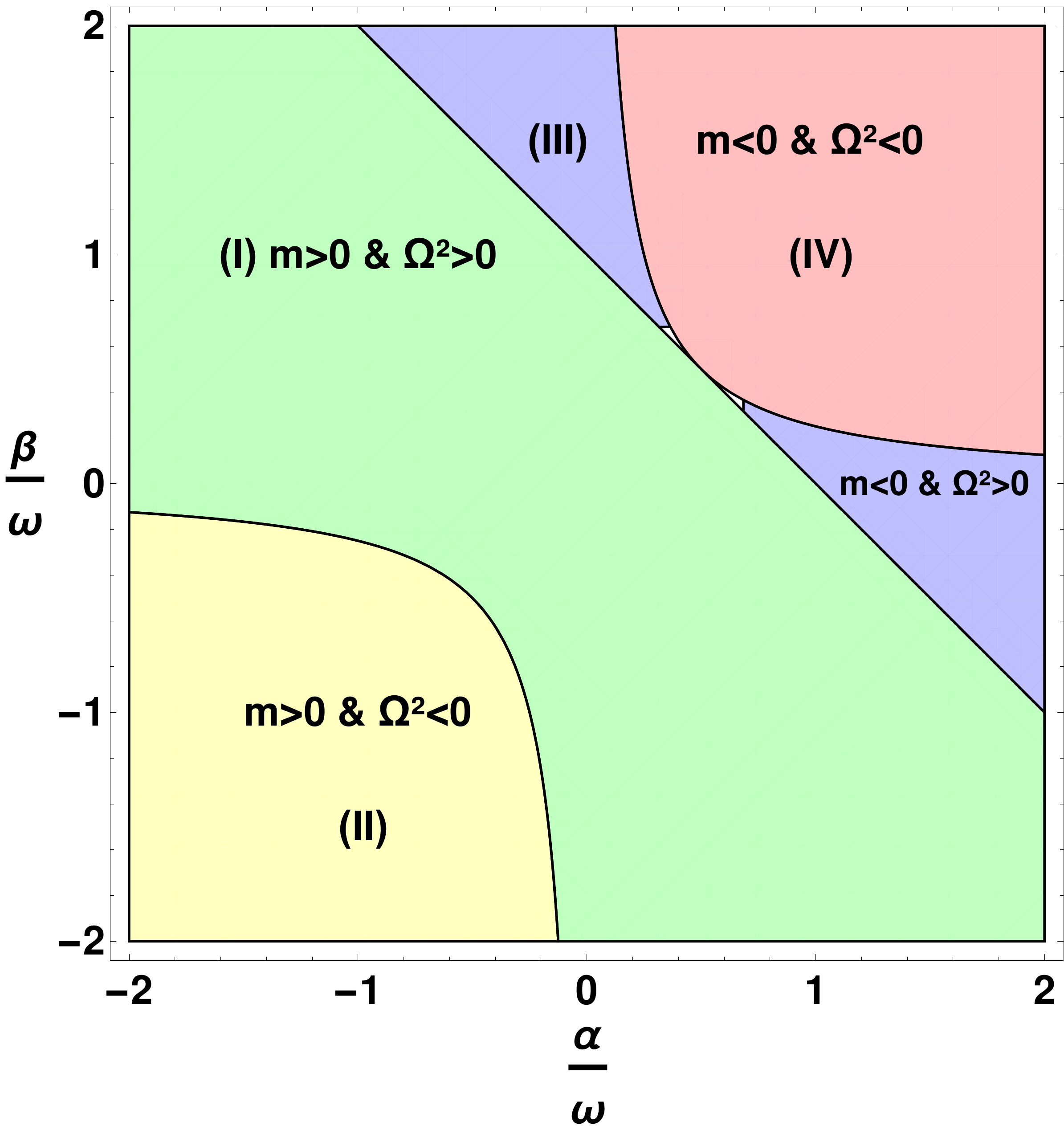}
\caption {Different regions of the parameter space in terms of the sign of $m(\omega,\alpha,\beta,b_0)$ and $\Omega^2(\omega,\alpha,\beta,b_0)$. }\label{fig:fig3}
\end{figure}

\subsection{Gelf'and Triplet.} \label{gelfand}
In general, neither $H$ nor $H_c$ have eigenfunctions in the Hilbert space ${\cal H}$. To overcome this problem, and to be able to compute the mean value of observables, we shall use the Gel'fand triplet \cite{gelfand,bhom}. Let us briefly review the essentials of the formalism.

To describe a quantum system we need a Hausdorff vector space with a convex topology and a scalar product, $(\Psi,\tau)$. The Hilbert space with the topology $\tau_H$, $({\mathcal{H}}, \tau_H)$, is the completion of $(\Psi,\tau_d)$. Let us define another completion of $(\Psi,\tau_d)$, with a finner topology $\tau_\Phi$, $(\Phi,\tau_\Phi)$, so that $\Phi \subset {\mathcal H} \Rightarrow {\mathcal H}^* \subset \Phi^*$. Here, 
$\Phi^*$ is the dual space of $\Phi$, 
$\Phi^*=\{ F | F: \Phi \rightarrow { \mathbb{C}}, F(v)=\langle F|v \rangle, ~v\in \Phi \}$. Also, we shall introduce the antidual space of $\Phi$, $\Phi^\times$, that is the space of the antilineal functionals on $\Phi$, 
$\Phi^\times =\{ G | G: \Phi \rightarrow { \mathbb{C}}, G(v)=\langle v|G \rangle, ~v\in \Phi  \}$. Along this lines, we obtain the Gel'fand triplet 

\beqn 
\Phi \subset {\cal H} \subset \Phi^\times.
\label{gelfand3}
\eeqn 
The extension of the hamiltonian operators $H$ and $H_c$ on ${\cal H}$ are the hamiltonian operators $H^\times$ and ${H_c}^\times$ on $\Phi^\times$, respectivelely. The  eigenfunctions of $H^\times$ and ${H_c}^\times$ are functionals of $\Phi^\times$.


In $\Phi^\times$, the stationary Schr\"odinger equation for $H$ and $H_c$ can be written as 

\beqn 
H^\times ~{\widetilde \phi}(x)= E~ {\widetilde \phi} (x), \label{eigeneq}
\eeqn
and

\beqn 
H_c^\times ~{\overline{\psi}}(x)= E_c~{\overline{\psi}} (x). \label{eigeneqc}
\eeqn
For $\om-\al-\be \neq 0$ we shall introduce the similarity transformation:
\beqn
\Upsilon =\re^{-\frac {\alpha-\beta}{\omega-\alpha-\beta}\frac { x^2} {2 b_0^2}},
\eeqn
so that

\beqn 
\Upsilon~ H^\times ~\Upsilon^{-1}&=& h^\times,
\nonumber \\
\Upsilon^{-1}~H_c^\times ~\Upsilon  &=& h^\times,
\label{simi}
\eeqn
with
\beqn
h^\times \phi(x)=-\frac{\hbar^2}{2 m }  \frac {{\rm d}^2 \phi(x)}{{\rm d} x^2}+\frac 1 2 k  x^2 \phi(x)= E ~\phi(x),
\label{hoscx}
\eeqn
the eigenfunctions of $H^\times$ and $H^\times_c$ are given by ${\widetilde \phi}(x)= \Upsilon^{-1} \phi(x)$ and ${\overline{\psi}}(x)= \Upsilon \phi(x)$, respectively.

Eq. (\ref{hoscx}) can be interpreted as the  Schr\"odinger equations corresponding to a potential of the form

\beqn
u(x)=\frac 12 k x^2= \frac 12 m \Om^2 x^2.
\label{ux}
\eeqn

From Eq.(\ref{simi}), it can be conclude that there is a similarity relation between $H^\times$ and $H^\times_c$, namely $U H^\times= H^\times_c U$, with $U=\Upsilon^2$.

As $H^\times$ is pseudo-hermitian, we can introduce a new inner product, $\langle . | . \rangle_U$, in terms of the positive define operator $U$

\beqn
\langle . | . \rangle_U: H^\times \times H^\times \rightarrow \mathbb C,~ \langle{\widetilde \phi}_\nu | {\widetilde \phi}_{\nu'}\rangle_U=\langle {\widetilde \phi}_\nu U | {\widetilde \phi}_{\nu'}\rangle_.
\eeqn 
It can be observed that the set $\{ {\overline \psi}_{\nu}\rangle, |{\widetilde \phi}_\nu\rangle \}$ is bi-orthogonal:

\beqn 
\langle {\widetilde \phi}_\nu | {\widetilde \phi}_{\nu'}\rangle_U & = & 
\langle {\overline{\psi}}_\nu | {\widetilde \phi}_{\nu'}\rangle =\langle \phi_\nu | \phi_{\nu'}\rangle=\delta_{\nu {\nu'}}.
\eeqn 
The identity operator can be written as

\beqn 
\mathds 1 =\sum_\nu | {\widetilde \phi}_\nu \rangle
\langle {\overline{\psi}}_\nu|.
\eeqn 

Given a pseudo-hermitian operator $\hat{Q}$, $\hat{Q}= \Upsilon q \Upsilon^{-1}$ with $q^\dagger = q$, its mean value can be computed as 

\beqn 
\langle {\widetilde \phi} | \hat{Q}|{\widetilde \phi} \rangle_U & = & 
\langle {\widetilde \phi} | U \hat{Q}|{\widetilde \phi} \rangle. 
\eeqn 

Thus, associated with the operators $p$ and $x$, we have

\beqn 
\hat{P}& = & \Upsilon p \Upsilon^{-1}=
{\hat p} +
\uni \hbar 
\frac {\alpha-\beta}{(\omega-\alpha-\beta)b_0^2}  {\hat x} , \nonumber \\
\hat{X}& = & \Upsilon x \Upsilon^{-1}= x,
\label{PX1}
\eeqn 
which is consistent with Eq.(\ref{PX}).

In what follows, we shall present the generalized eigenfunctions and the corresponding spectrum in the different regions.

\subsection{Spectrum and Generalized Eigenfunctions.}\label{regions}

\subsubsection{Region I.}

In the PT-symmetry phase, Eq. (\ref{hoscx}) reduces to the usual harmonic oscillator. Consequently, the spectrum and the eigenfunctions are given by

\beqn
E_n = E_{cn} & = & \hbar \sqrt{\om^2 - 4 \al \be}~\left( n+\frac 12 \right), \nonumber \\
\widetilde{\phi}_n(x) & = & 
{\rm e^{\frac {\alpha-\beta}{\omega-\alpha-\beta} \frac { x^2} {2 b_0^2}}}
\frac{2^{-\frac n 2}}{\sqrt{\pi} n!} \frac {\sigma}{b_0  }~ {\rm e}^{-\frac{x^2}{2 b^2_0} \sigma^2}~
H_n \left ( \frac{x}{b_0} \sigma \right), \nonumber \\
\overline{\psi}_n(x) & = & 
{\rm e^{-\frac {\alpha-\beta}{\omega-\alpha-\beta} \frac { x^2} {2 b_0^2}}}
\frac{2^{-\frac n 2}}{\sqrt{\pi} n!} \frac {\sigma}{b_0  }~ {\rm e}^{-\frac{x^2}{2 b^2_0} \sigma^2}~
H_n \left ( \frac{x}{b_0} \sigma \right), \nonumber \\
\eeqn
with $\sigma= \sqrt{ \frac {m \Om}{\hbar} }b_0$ and $\Om=\sqrt{\om^2 - 4 \al \be}$.

In terms of the eigenfunctions of $H^\times$ and ${H_c}^\times$,  the bi-orthogonality relation is given by
\beqn
\int_{-\infty}^\infty {\overline{\psi}_m^*(x)} \widetilde{\phi}_n(x)\ dx= \delta_{nm}.
\eeqn
and the completeness relation is 

\beqn
\sum_{n}^\infty \overline{\psi}^*_n(x) \widetilde{\phi}_m(x')=\delta(x-x').
\eeqn

\subsubsection{Region III.}

In this region, the parameter $m$ takes negative values. As it is proved in Appendix A, we can define $\sg=\sqrt{ \frac {|m \Om|}{\hbar} }b_0 \in  \mathbb{R}$, as in the previous case. So that 

\beqn
E_n = E_{c n} & = & -\hbar \sqrt{\om^2 - 4 \al \be}~\left( n+\frac 12 \right), \nonumber \\
\widetilde{\phi}_n(x) & = & \tfrac{2^{-\frac n 2}}{\sqrt{\pi} n!} \tfrac {\sigma}{b_0  }~ 
{\rm e^{\frac {\alpha-\beta}{\omega-\alpha-\beta} \frac { x^2} {2 b_0^2}}}{\rm e}^{-\frac{x^2}{2 b^2_0} \sigma^2}~
H_n \left ( \frac{x}{b_0} \sigma \right), \nonumber \\
\overline{\psi}_n(x) & = & \tfrac{2^{-\frac n 2}}{\sqrt{\pi} n!} \tfrac {\sigma}{b_0  }~ 
{\rm e^{-\frac {\alpha-\beta}{\omega-\alpha-\beta} \frac { x^2} {2 b_0^2}}}{\rm e}^{-\frac{x^2}{2 b^2_0} \sigma^2}~
H_n \left ( \frac{x}{b_0} \sigma \right). \nonumber \\
\eeqn

The bi-orthogonality and completeness relation is the same as in the previous case.

\subsubsection{Region II.}

In this region, $m>0$ and $\Om^2<0$. The potential $u(x)$ of Eq.(\ref{ux}) corresponds to that of a parabolic barrier \cite{chu1,chu2,maru,bermudez}. Both hamiltonians, $H^\times$ and $H_c^\times$, display continuous spectrum as well as resonant and anti-resonant discrete states. 

In Appendix A we present the construction of generalized eigenfunctions and the corresponding eigenvalues. Let us summarize the results as follows.

The generalized eingenfunctions of $H^\times$, $\wid{\phi}_n^\pm(x)$, with eigenvalues $E_n^\pm=\pm \uni \hbar |\Om| \left( n +\frac 12 \right)$ are given by
\beqn 
\wid{\phi}_n^\pm(x) & = & 
{\rm e^{\frac {\alpha-\beta}{\omega-\alpha-\beta} \frac { x^2} {2 b_0^2}}}
\phi_n^\pm  \left( x \right),
\eeqn 
while, the generalized eigenfunctions of $H_c^\times$, $\ovl{\psi}_n^\pm(x)$, with eigenvalues $E_{cn}^\pm=\mp \uni \hbar |\Om| \left( n +\frac 12 \right)$ are given by
\beqn 
\ovl{\psi}_n^\pm(x) & = & 
{\rm e^{-\frac {\alpha-\beta}{\omega-\alpha-\beta} \frac { x^2} {2 b_0^2}}}
\phi_n^\mp \left( x \right).
\eeqn  
 and 
\beqn
\phi_n^-(x) & = &  
\sqrt { \tfrac{\sqrt{i} \sg } {b_0} \frac{1} {2^n n!}}
\re^{- \uni \sg^2 \frac {x^2} {2 b_0^2}}
H_n\left( \frac {\sqrt{ \uni} \sg}{b_0} x \right),\nonumber \\
\phi_n^+(x)    & = & {\phi_n^-(x) }^*,
\eeqn
with $\sg= \sqrt{\frac{|m \Om|}{\hbar} }b_0 $.

The bi-orthogonality relation reads

\beqn
\int_{-\infty}^\infty (\ovl{\psi}_m^\pm(x))^* \wid{\phi}_n^\pm(x)= \delta_{nm},
\eeqn

and the completeness relation is given by
\beqn
\sum_{
\begin{array}{l}
n=0\\
s = \pm
\end{array}
}^\infty~(\overline{\psi}_n^s(x))^* \wid{\phi}_n^s(x')& = & 
\delta(x-x').
\eeqn

The generalized eigenfunctions associated to the continuous spectrum, $E \in (-\infty, +\infty)$, are given by

\beqn
\wid \phi_\pm^E(x) & = & {\rm e^{\frac {\alpha-\beta}{\omega-\alpha-\beta} \frac { x^2} {2 b_0^2}}}~
\phi_\pm^E(x), \\
\ovl \psi_\pm^E(x) & = & {\rm e^{-\frac {\alpha-\beta}{\omega-\alpha-\beta} \frac { x^2} {2 b_0^2}}}~
\phi_\mp^E(x),
\eeqn
with $\nu= -\uni \frac{E}{\hbar |\Om|}-\frac 12$ and 

\beqn
\phi_\pm^E(x) & = & \mathcal{C}  ~
\Gamma(\nu+1) D_{-\nu-1} \left (\mp \sqrt{2\uni}  \sigma \frac{x}{b_0} \right).
\eeqn

The bi-orthogonality and the completeness relation can be written as

\beqn
\int_{-\infty}^\infty ~(\overline{\psi}_\pm^E(x))^*
\wid \phi_\pm^{E'}(x) {\rm dx}= \delta(E-E'),\nonumber \\
\sum_{s=\pm}\int_{-\infty}^\infty ~(\overline{\psi}_s^E(x))^*
\wid \phi_s^{E}(x) {\rm dE}= \delta(x-x').\nonumber \\
\eeqn

To complete the analysis of this region we have to discuss the analytical properties of the previous solutions.

It is easy to see that the poles of $\phi_\pm^E(x)$ are those  of $\Gamma(\nu+1)$, that is: $-n=\nu+1$, with $n \in \mathbb N$ and $E_n=\uni \hbar |\Om|\left( n+ \frac 12 \right)$. 

In the other hand, the poles of $(\phi_\pm^E(x))^*$ are those  of $\Gamma(-\nu)$, that is: $n=\nu$, with $n \in \mathbb N$ and $E_n=-\uni \hbar |\Om|\left( n+ \frac 12 \right)$. As shown in Appendix A, we shall introduce $\eta^E_\pm(x)=(\phi_\pm^E(x))^*$, which are eigenfunctions of $h^\times$ by replacing $E \leftrightarrow -E$.

In the coordinate representation, we have 
$\phi_\pm^E(x)= \langle x |\phi_\pm^E\rangle$ and $\eta_\pm^E(x)= \langle x | \eta_\pm^E\rangle$. Consequently: 

\beqn 
|x \rangle = \mathcal C \sum_{s=\pm } \inte
\left( \eta_s^E(x) |\phi_s^E \rangle+ \phi_s^E(x) |\eta_s^E \rangle \right){\rm {dE}},
\eeqn 
so that $\langle x| x' \rangle=\delta(x-x')$.

A function $\xi(x)= \langle \zeta | x \rangle$, can be written as

\beqn
\xi(x)& = & \xi^+(x)+ \xi^-(x), \nonumber \\ 
\xi^+(x) & = & \mathcal C \sum_{s=\pm } \inte
 \eta_s^E(x) \langle \zeta |\phi_s^E \rangle {\rm {dE}}, \nonumber \\
\xi^-(x) & = & \mathcal C \sum_{s=\pm } \inte
 \phi_s^E(x) \langle \zeta |\eta_s^E \rangle {\rm {dE}},
\label{fx}
\eeqn


Let us take $\Phi$ of Eq.(\ref{gelfand3}) as the space of Hardy class function \cite{gadella1,gadella2}. We shall define 
$\mathcal H_+$ as the Hardy class functions in the upper half-plane $\mathbb C^+$, that is the set of the analytic functions, $f(z)$, in $\mathbb C^+$ so that 

$$ \int_{-\infty}^\infty |f(x+\uni y)|^2 dx <\infty.$$
In the same way, $\mathcal H_-$ is the set of the Hardy class functions in the lower half-plane, $\mathbb C^-$.

It should be noticed that an $\mathcal H_\pm$ function is completely determined by its value on $\mathbb R$. We shall define

\beqn 
\Xi_- & = & \{ \xi \in \Phi | f(E)= \langle \xi |\eta^E_\pm \rangle \in \mathcal H_- \}, \nonumber \\
\Xi_+ & = & \{ \xi \in \Phi | f(E)= \langle \xi |\phi^E_\pm \rangle \in \mathcal H_+ \}. 
\eeqn 
Following the lines of \cite{chu2,maru}, we can expand functions $\xi^-(x) \in \Xi^-$ and $\xi^+(x) \in \Xi^+$ as

\beqn 
\xi^-(x) & = & 
\sum_\pm \int_{-\infty}^\infty {\rm d E} ~ 
\phi_\pm^E(x) \langle \phi_\pm^E |\xi^-\rangle= 
\sum_{n=0}^\infty \phi_n^-(x) \langle \phi_n^+|\xi^-\rangle, \nonumber\\
\xi^+(x) & = & \sum_\pm \int_{-\infty}^\infty {\rm d E} ~ \eta_\pm^E(x)
\langle \eta_\pm^E |\xi^+\rangle= 
\sum_{n=0}^\infty \phi_n^+(x) \langle \phi_n^-|\xi^+\rangle, \nonumber\\
\eeqn
respectively. In the same manner, we can construct the 
following spectral resolution for $h^\times$:

\beqn 
h^\times= \sum_\pm \int_{-\infty}^\infty {\rm d E} E | \phi_\pm^E \rangle
\langle \phi_\pm^E |= \sum_{n=0}^\infty E_n^- |\phi_n^-\rangle \langle \phi_n^+|, 
\eeqn
on $\Xi_-$, and 

\beqn 
h^\times= \sum_\pm \int_{-\infty}^\infty {\rm d E} E | \eta_\pm^E \rangle
\langle \eta_\pm^E |= \sum_{n=0}^\infty E_n^+ |\phi_n^+\rangle \langle \phi_n^-|, 
\eeqn
on $\Xi_+$. 

Consequently, $H^\times= \Upsilon^{-1} h^\times \Upsilon$ and 
$H_c^\times= \Upsilon h^\times \Upsilon^{-1}$:

\beqn 
H^\times & = & 
\sum_\pm \int_{-\infty}^\infty {\rm d E} ~E ~\left( | \wid \phi_\pm^E \rangle \langle \ovl \phi_\pm^E | +
|\wid  \eta_\pm^E \rangle \langle \ovl \eta_\pm^E | \right) \nonumber \\
& = & \sum_{n=0}^\infty \left( E_n^- |\wid \phi_n^-\rangle \langle \ovl \phi_n^+|
+ E_n^+ |\wid \phi_n^+\rangle \langle \ovl \phi_n^-| \right), 
\eeqn
and 
\beqn 
H_c^\times & = & 
\sum_\pm \int_{-\infty}^\infty {\rm d E} ~E ~\left( 
| \ovl \phi_\pm^E \rangle \langle \wid \phi_\pm^E | +
| \ovl \eta_\pm^E \rangle \langle \wid \eta_\pm^E | 
\right) \nonumber \\
& = & \sum_{n=0}^\infty \left( 
  E_n^- |\ovl \phi_n^-\rangle \langle \wid \phi_n^+|
+ E_n^+ |\ovl \phi_n^+\rangle \langle \wid \phi_n^-| \right). 
\eeqn

\subsubsection{Region IV.}

In Region IV, $m<0$ and $\Om^2<0$. The results are similar to the ones of Region II, $E \leftrightarrow -E$. See Appendix A.

\subsubsection{Boundary I-II and III-IV.}

In both boundaries, I-II and III-IV, $\Om$ takes the value $\Om=0$.
When $\Omega=0$ and $\omega -(\alpha+\beta) \ne 0$, the problem reduces to that of a free particle of energy $E$. As shown in Appendix A, the generalized eigenfunctions can be written as 

\beqn 
\wid \phi(x) & = & \left( A {\rm e}^{i k x}+B {\rm e}^{-i k x} \right ) 
\re^{-\frac{\om + 2 \be }{\om-2 \be} \frac{x^2}{2 b_0^2}}, \nonumber\\ 
\ovl \psi(x) & = & \left( A {\rm e}^{i k x}+B {\rm e}^{-i k x} \right ) 
\re^{\frac{\om + 2 \be }{\om-2 \be} \frac{x^2}{2 b_0^2}},
\eeqn 
with 
$k= \sqrt{\frac{2 E}{\hbar (\omega-\alpha-\beta) b_0^2}}$.

\subsubsection{Exceptional Points.}
As pointed out in \cite{fring-sw3}, at the boundary I-II and III-IV we observe the presence of EPs. 
At these points, the discrete eigenvalues and the eigenfunctions of region I and II, and of region III and IV, are coalescent. 
At each EP the energy of the state converges to $E=0$ for all values of $n$, and from both sides of the boundary the eigenfunctions converge to 

\beqn
\widetilde{\phi}(x) & = & \left(c_1 x+c_0\right) e^{-\frac{x^2 (w+2 \beta)}{2 (w-2 \beta )}}, \nonumber \\
\overline{\psi}(x)  & = & \left(d_1 x+d_0\right) e^{~\frac{x^2 (w+2 \beta)}{2 (w-2 \beta )}},
\nonumber \\
E & = & 0.
\eeqn

Thus at the boundary of Regions I-II and III-IV, the spectrum consists of PEs of infinite order \cite{mannheim1,mannheim2,bookcmb,smilga}, with $E=0$, which resides within the continuum spectrum \cite{bersch}.
The details are given in Appendix C.

\subsubsection{Boundary I-III.}

To study the boundary between Regions I and III, we have to look at the Hamiltonian of Eq.(\ref{hxp}). If $\omega-(\alpha+\beta)=0$, it reads

\beqn
H^\times(\theta) & = & 
\hbar (\alpha +\beta ) \left( \frac {{\hat x}} {b_0}\right)^2 \nonumber \\
& &+  \hbar \frac{(\alpha-\beta)}{2} \left( 2~\hat{x} \frac{ {\bf i}}{\hbar} \hat{p}+1 \right),
\eeqn

and its adjoint is given by

\beqn
H_c^\times(\theta) & = & 
\hbar (\alpha +\beta ) \left( \frac {x} {b_0}\right)^2 \nonumber \\
& &+  \hbar \frac{(\beta-\alpha)}{2} \left( 2~\hat{x} \frac{ {\bf i}}{\hbar} \hat{p}+1 \right).
\eeqn

We shall introduce a new similarity transformation by defining the operator
\beqn
\tau=\re^{\frac {\al+\be}{\al-\be} \frac{x^2}{2 b_0^2}}, 
\eeqn
it results 

\beqn 
\tau~ H^\times ~\tau^{-1}&=& h^\times,~~~
\tau^{-1}~H_c^\times ~\tau  =-h^\times,
\label{simi}
\eeqn
with
\beqn
h^\times \phi(x)=\frac{\hbar (\al -\be)}{2}\left(2 x {\rm{\frac d {dx}}}+ 1 \right) \phi(x)= E \phi(x).
\label{h13}
\eeqn
It is straightforward to see that $H^\times$ and $H_c^\times$ are anti-pseudo-hermitian \cite{antipseudo} at the boundary, that is $H^\times=-S^{-1} H_c^\times S$, with $S= \tau^2$.

The spectrum of hamiltonian of Eq.(\ref{h13}) is real. The generalized eigenfunctions of $H^\times$ and $H^\times_c$ are given by ${\widetilde \phi}(x)= \tau^{-1} \phi(x)$ and ${\overline{\psi}}(x)= \tau\phi(x)$, respectively.

The generalized eigenfunctions of $H^\times$ are given by

\beqn
\widetilde{\phi}_n^+(x) & = &  \frac{1}{\sqrt{n!}}      \re^{-\frac {\al+\be}{\al-\be} \frac{x^2}{2 b_0^2}} x^n, 
~~ \wid E_n^+= E_n, \nonumber \\
\widetilde{\phi}_n^-(x) & = &  \frac{(-1)^n}{\sqrt{n!}} \re^{-\frac {\al+\be}{\al-\be} \frac{x^2}{2 b_0^2}} \delta^{(n)}(x),
~~ \wid E_n^-= -E_n. \nonumber \\
\eeqn

While for  $H_c^\times$ the corresponding generalized eigenfunctions are:

\beqn
\overline{\psi}_n^+(x) & = &  \frac{1}{\sqrt{n!}}      \re^{\frac {\al+\be}{\al-\be} \frac{x^2}{2 b_0^2}} x^n,
~~ \ovl E_n^+= -E_n \nonumber\\
\overline{\psi}_n^-(x) & = &  \frac{(-1)^n}{\sqrt{n!}} \re^{\frac {\al+\be}{\al-\be} \frac{x^2}{2 b_0^2}} \delta^{(n)}(x),
~~ \ovl E_n^-= E_n. \nonumber\\
\eeqn
with $E_n=\hbar (\al-\be) \left ( n+ \frac 12 \right)$.

Thus, the bi-orthogonality relations are given by

$$ \int_{-\infty}^\infty (\overline{\psi}_m^\mp(x))^* \overline{\phi}_n^\pm(x)~ dx= \delta_{nm}.$$

The details are presented in Appendix A.

The eigenfunctions with positive spectrum of the boundary between regions I-III can be obtained by a limit procedure \cite{mannheim1,mannheim2} from the eigenfunctions of region I. In the same form, the eigenfunctions corresponding to negative values of the spectrum can be obtained from the eigenfunctions of Region III. The details are presented in Appendix B.

\section{Time evolution of Observables.} \label{evolution}

Let us discuss in first place regions I and III. In these cases, the spectrum of $H^\times$ is real and discrete. It is not difficult to prove that the mean value of the operators $\hat X$ and $\hat P$ of Eq.(\ref{PX1}) between states different $|\wid \phi_m\rangle$ obey 

\beqn
\langle {\wid \phi}_m | U {\hat X} |{\wid \phi}_n\rangle & = & 
\int_{-\infty}^{\infty}~{{\ovl \psi}^\pm}^*_n(x){\hat X} {\wid \phi}_n(x) {\rm dx} \nonumber \\
& = &
\frac {b_0}{\sqrt {2}}
\left( \sqrt {n+1} \delta_{m,n+1}+  \sqrt {n} \delta_{m,n-1} \right), \nonumber\\
\langle{\wid \phi}_m| U {\hat X}^2 |{\wid \phi}_n
\rangle & = & 
\int_{-\infty}^{\infty}~{{\ovl \psi}^\pm}^*_n(x){\hat X}^2 {\wid \phi}_n(x) {\rm dx} \nonumber \\
& = &
\frac {b_0^2}{2}
\left( \sqrt {(n+2)(n+1)} \delta_{m,n+2}  \right .\nonumber \\
& &         ~~~~~~~~~+(2 n +1) \delta_{m,n} \nonumber \\
& & \left . ~~~~~~~~~+\sqrt {n(n-1)} \delta_{m,n-2} \right). \nonumber\\
\langle{\wid \phi}_m| U {\hat P} |{\wid \phi}_n\rangle & = & 
\int_{-\infty}^{\infty}~{{\ovl \psi}^\pm}^*_n(x){\hat P} {\wid \phi}_n(x) {\rm dx} \nonumber \\
& = &
\frac {{\bf i} \hbar}{\sqrt {2}b_0}
\left( \sqrt {n+1} \delta_{m,n+1}- \sqrt {n} \delta_{m,n-1} \right), \nonumber\\
\langle{\wid \phi}_m| U {\hat P}^2 |{\wid \phi}_n\rangle & = & 
\int_{-\infty}^{\infty}~{{\ovl \psi}^\pm}^*_n(x){\hat P}^2 {\wid \phi}_n(x) {\rm dx} \nonumber \\
& = &
-\frac {\hbar^2} {2 b_0^2}
\left( \sqrt {(n+2)(n+1)} \delta_{m,n+2}  \right .\nonumber \\
& &         ~~~~~~~~~-(2 n +1) \delta_{m,n} \nonumber \\
& & \left . ~~~~~~~~~+\sqrt {n(n-1)} \delta_{m,n-2} \right). \nonumber\\
\eeqn

The time evolution of a given initial state $| I(0) \rangle$,  $| I(0) = \sum_k c_k | \wid \phi_k \rangle$, such that $\langle I(0)| I(0) \rangle_S=1$, is given by

\beqn 
\langle I (t) |\hat O | I(t)\rangle_U & = &  
\langle I (0) |\re^{-\uni H^\dagger t}S \hat O \re^{-\uni H t} | I(0)\rangle
\nonumber \\
& = & \sum_{nm} c_n c_m^* \re^{\uni (E_m-E_n)t}
\langle {\widetilde \phi}_m | \Upsilon \hat{o} \Upsilon^{-1}|{\widetilde \phi}_n \rangle.
\nonumber \\
\eeqn 

In regions II and IV, the spectrum of $H^\times$ consists of real continuous eigenvalues and discrete resonant and anti-resonant states.

If $U(t)=\re^{-\uni H t}$ is the operator for the time evolution in $\mathcal H$, $U(t)^\times=\re^{i H^\times t}$ is the operator for the time evolution in $\mathcal H^\times$ \cite{maru}. Thus $U(t)^\times |\phi_n^\pm \rangle= 
\re^{\mp \hbar |\Om| (n+\frac 12) t}|\phi_n^\pm \rangle$, and a wave function will evolve in time under the action of $H^\times$ as

\beqn 
\xi(x,t) &= & \xi^+(x,t)+ \xi^-(x,t), \nonumber \\
\xi^-(x,t) & = &  
\sum_{n=0}^\infty \re^{~\hbar |\Om| (n+\frac 12) t} \wid \phi_n^-(x) \langle  \ovl \phi_n^+|\xi^-\rangle, \nonumber\\
\xi^+(x,t) & = &  
\sum_{n=0}^\infty \re^{ -\hbar |\Om| (n+\frac 12) t} \wid \phi_n^+(x) \langle \ovl \phi_n^-|\xi^+\rangle.
\eeqn
Given a particular problem \cite{scat-znojil,scat-ahmed,scat-mosta1,scat-mosta2}, we may have to consider only one of the contributions to $\xi(x,t)$, and take the other as a background \cite{maru}, or both of them if we model a system with gain-loss balance. 


\section{Conclusions.}
\label{conclu}

In this work, we have studied the non-hermitian Swanson hamiltonian, both in the PT-symmetry and in the non-PT symmetry phase. 
As a result, we have mapped the Swanson model to different physical systems depending on the adopted values for the parameters $\al/om$ and $\be/\om$. We have classified regions and their boundaries. 
We have shown that Region I corresponds to the usual harmonic oscillator, Region III to a harmonic oscillator with negative mass, Region II represents a parabolic barrier, and region IV a parabolic barrier for a particle with negative mass.
We have used the formalism of the Gel'fand triplet to construct the generalized eigenfunctions and the corresponding spectrum in each region.
We have shown that it is possible to construct metric operators in the Rigged Hilbert Space. Also, we have proved that we can establish a bi-orthogonality relation among the generalized functions of $H^\times$ and $H_c^\times$.  
In the same line, we have formalized the computation of
the mean value of observables and the time evolution of the system in the different regions of the model space. 
 Also, we have verified that the boundary between the regions I-II and III-IV is formed by Exceptional Points of infinite order embedded in a continuum spectrum \cite{mannheim1,mannheim2}. An interesting feature results from the study of the boundary between the regions I and III, $H^\times$ and of $H_c^\times$ are anti-pseudo-hermitian \cite{antipseudo}. \\
Work is in progress concerning the computation of the evolution of different initial states as a function of time for different regions of the model space. Particularly, in the boundary between Regions I-III where $H^\times$ and $H_c^\times$ are anti-pseudo-hermitian \cite{antipseudo}, and between Regions I-II and III-IV, where the continuum spectrum includes the presence of Exceptional Points with energy $E=0$ \cite{garmon3,garmon4}.  

\section*{Appendix A}
We shall follow the works of \cite{chu1,chu2} and \cite{maru}
to construct the generalized eigenfunctions in the Rigged Hilbert Space. Let us briefly review the essentials of the procedure.

After performing the similarity transformations of Eqs. (\ref{simi}), for $\omega - \alpha - \beta\neq 0$, the eigenvalue problem for both hamiltonians, $H^\times$ and its adjoint $H^\times_c$, can be reduce to find the spectrum and the generalized eigenfunctions of

\beqn
h^\times \phi(x)=-\tfrac{\hbar^2}{2 m }  \tfrac {{\rm d}^2 \phi(x)}{{\rm d} x^2}+\tfrac 1 2 m \Om^2  x^2 \phi(x)= E ~\phi(x).
\label{hamh}
\eeqn

\subsection*{Regions I and III.}

In Regions I and III, the parameter $\Omega$ takes real values, $\Omega^2 >0$. 
In Region I, the parameter $m$ takes positive values, $m=|m|$, while in Region III we have $m=-|m|$. Taken this fact into account, we can write the previous equation as  

\beqn
-\frac{\hbar}{2 |m| \Om }  \frac {{\rm d}^2 \phi(x)}{{\rm d} x^2}+\frac 1 2 \frac{ |m| \Om}{\hbar}  x^2 \phi(x)= \frac{\varepsilon}{\hbar \Om}~\phi(x), \nonumber \\
\eeqn
with $\varepsilon=E$ in Region I, and $\varepsilon=-E$ in Region III.

Introducing the variables ${\hat z}= \sqrt{\frac{ |m| \Om}{\hbar}} x$ and ${\hat p}_z= -\uni  \frac{{\rm d~~}}{{\rm d} z}$, the eigenvalue problem is given by

\beqn
\frac 12 \left( {\hat p}_z^{2}+{\hat z}^{2} \right) \phi(z)= \frac{\varepsilon}{\hbar \Om} \phi(z).
\label{hx}
\eeqn

If $\Om= |\Om|$ the parameter $\sg \in \mathbb R$, in this case  let us introduce the following representation in terms of the new  operators $\hud$ and $\hu$ :
 
\beqn
{\hat u^\dagger} & =& \frac{{\hat z} - \uni {\hat p}_{z}}{\sqrt{2}},\nonumber \\
{\hat u} & =& \frac{{\hat z} + \uni {\hat p}_{z}}{\sqrt{2}},
\label{uv}
\eeqn
with $[u,u^\dagger]=1$. In terms of ${\hat u}^\dagger$ and $\hat{u}$, the hamiltonian of Eq.(\ref{hx}) can be written in the well known form:  

\beqn
h^\times |\phi\rangle=  \hbar |\Omega|\left ({\hat u}^\dagger {\hat u} +\frac 12 \right)|\phi\rangle=
\varepsilon |\phi\rangle.
\label{hdist}
\eeqn

As it is well known \cite{messiah}, in this representation the eigenfunctions and eigenvalues correspond to the discrete solutions of Eq.(\ref{hdist}).  The operator $\hud \hu$ is an hermitian positive define operator, its lower eigenvalue takes the $0$ value. It corresponds to a state of energy $\frac {\hbar \Om}{2}$ \cite{messiah}:  
\beqn
{\hat u} |\phi_0\rangle =0, ~~\epsilon_0=\frac 12 \hbar |\Om|.\\
\eeqn
The other states can be constructed as usual:

\beqn
|\phi_n\rangle = \frac{1}{\sqrt{n!}}{\hat u}^{\dagger n}|\phi_0\rangle, ~~ \varepsilon_n = \hbar |\Om| \left( n +\frac 12\right).
\eeqn
The set of eigenvectors $\{ | \phi_n \rangle \}$ is a complete and orthogonal set, that is:

\beqn 
\mathds 1 =\sum_{n=0}^\infty | \phi_n \rangle \langle \phi_n |, ~~
\langle \phi_m | \phi_n \rangle = \delta_{m,n}. \nonumber \\
\eeqn 

Let us briefly reviewed the
construction of the representation in terms of functions of $z$, we shall use the generating function \cite{messiah}. The generating function satisfies the following property 

\beqn 
G(t,z)= \sum_{n} c_n \phi_n(z) \frac{t^n}{\sqrt { n!}}. 
\eeqn
Consider as function of $z$, and taking $c_n=n!^{-1/2}$, it represent the vector 
$$\sum_n  \frac {(\hud t)^n}{n!} |0 \rangle,$$
so that $G(t,z)$ can be obtained as $G(t,z)=\langle z | \re^{\hud t} | \phi_0 \rangle$.
For a complete development, the reader is referred to \cite{messiah}. By noticing that
$\re^{\hud t}$ can be rewritten in terms of $z$ and $p_z$, and that $\phi_0(z)= \pi^{-1/4} \re^{-z^2/2}$ is the solution of $\left( z+ \frac{\rm d~} {\rm{dz}} \right ) \phi_0(z)=0$, we obtain 

\beqn 
G(t,z) & = & \pi^{-1/4}
\re^{-\frac {z^2}2 + 2 z \frac {t}{\sqrt 2} -\left ( \frac {t}{\sqrt 2} \right)^2} \nonumber \\
& = & \sum_n ~\re^{-\frac {z^2}2} \frac {H_n(z)}{\sqrt{2 ^n n! \sqrt{\pi}}}\frac{t^n}{\sqrt {n!}}, 
\eeqn 
where we have used that

\beqn 
\re^{ 2 z \zeta -\zeta^2}= \sum_n ~H_n(z)\frac{\zeta^n}{ n!}.
\label{hermite}
\eeqn 
Consequently, $\phi_n(z)= \re^{-\frac {z^2}2} \frac {H_n(z)}{\sqrt{2 ^n n! \sqrt{\pi}}}$.

In this representation, the relations of completeness and orthogonality are given by:

\beqn
\sum_{n=0}^\infty \phi_n(z) \phi_n(z')& = &\delta(z-z'), \nonumber \\
\int_{-\infty}^\infty \phi_n(z) \phi_m(z) {\rm {dz}}&=&\delta_{nm}.
\eeqn 

In the representation of coordinates 

\begin{equation}
|z \rangle = (2 \pi)^{1/4} \sum_n \phi_n(z) |\phi_n\rangle,
\end{equation}
with 

\beqn 
\mathds 1 =\int_{-\infty}^\infty | z \rangle \langle z | {\rm {dz}}, ~~
\langle z | z' \rangle = \delta(z-z'). 
\eeqn 
To work in the representation of momentum, we have to perform a Fourier Transform

\beqn 
\phi(p_z) = \mathcal F [\phi_n(z)](p_z)= \frac 1{\sqrt {2 \pi} } 
\int_{-\infty}^\infty \phi_n(z) \re^{\uni p_z z} {\rm {dz} },
\eeqn 
also 

\beqn 
|p_z \rangle =  (2 \pi)^{1/4} \sum_n \uni^n \phi_n(p_z) |\phi_n\rangle,
\eeqn 
and 

\beqn 
\mathds 1 =\int_{-\infty}^\infty | p_z \rangle \langle p_z | {\rm {dz}}, ~~
\langle p_z | p_z' \rangle = \delta(p_z-p_z'). 
\eeqn 

\subsection*{Boundary I-III.}

In the boundary between Regions I and III, we have to find the generalized eigenfunctions of $h^\times$
\beqn
h^\times \phi(x)=\frac{\hbar (\al -\be)}{2}\left(2 x {\rm{\frac d {dx}}}+ 1 \right) \phi(x)= E \phi(x),
\label{h13}
\eeqn
It is straightforward to verify that the solutions are 

\beqn
\phi_n^+(x) =  &\frac{x^n}{\sqrt{n!}}, ~~ & E_n = \hbar (\al-\be) \left( n +\frac 12\right), \nonumber \\
\phi_n^-(x) =  &(-1)^n \frac{\delta^{(n)}(x)}{\sqrt{n!}}, ~~ & E_n = -\hbar (\al-\be) \left( n +\frac 12\right). \nonumber \\
\eeqn

\subsection*{Regions II and IV.}

In Regions II and IV, we have $\Omega^2<0$. 
Furthermore, in Region II $m=|m|$, while in Region IV we have $m=-|m|$. Taken this fact into account, we can write the previous equation as  

\beqn
-\frac{\hbar}{2 |m| \Om }  \frac {{\rm d}^2 \phi(x)}{{\rm d} x^2}+\frac 1 2 \frac{ |m| \Om}{\hbar}  x^2 \phi(x)= \frac{\varepsilon}{\hbar \Om}~\phi(x), \nonumber \\
\label{h24}
\eeqn
with $\varepsilon=E$ in Region II, and $\varepsilon=-E$ in Region IV.

Let us introduce the parameter $\sg= \sqrt{|m| \Om /\hbar}$, with $\Om=\pm \uni |\Om|$. We shall take $z=\re^{\uni \frac \pi 4} |\sg| x$, and define the new set of operators

\beqn
{\hat u^\dagger} & =& \frac{{\hat z} - \uni {\hat p}_{z}}{\sqrt{2}},\nonumber \\
{\hat v} & =& \frac{{\hat z} + \uni {\hat p}_{z}}{\sqrt{2}},
\label{uv2}
\eeqn
with $ [\hv,\hud]=1 $. To study the spectrum of the hamiltonian of Eq.(\ref{h24}),  we shall split it as follows:

\beqn 
 \uni \hbar |\Om|\left ( \hud \hv + \frac 12 \right ) |\phi^+ \rangle & = & \varepsilon |\phi^+ \rangle , 
\label{hpos}
\eeqn

\beqn
-\uni \hbar |\Om| \left ( \hvd \hu + \frac 12 \right ) |\phi^- \rangle & = & \varepsilon |\phi^- \rangle.
\label{hneg}
\eeqn 
We can proceed as before. It results 

\beqn 
\begin{array}{cl}
     \hv |\phi_0^+ \rangle =0, & \varepsilon_0^+=\uni   \frac{\hbar |\Om|} 2 \\
     |\phi_n^+ \rangle=\frac {{\hat u}^{\dagger n}}{\sqrt{n!}} |\phi_0^+ \rangle, & \varepsilon_n^+=\uni   \frac{\hbar |\Om|} 2 [n],
\end{array}
\eeqn 
and
\beqn 
\begin{array}{cl}
     \hu |\phi_0^- \rangle =0, & \varepsilon_0^-=-\uni   \frac{\hbar |\Om|} 2 \\
     |\phi_n^- \rangle=\frac {{\hat v}^{\dagger n}}{\sqrt{n!}} |\phi_0^- \rangle, & \varepsilon_n^-=-\uni   \frac{\hbar |\Om|} 2 [n].
\end{array}
\eeqn 
It is straightforward to prove that

\beqn
\langle \phi_m^\mp | \phi_n^\pm \rangle= \delta_{mn},
\eeqn 
and 


The eigenfunctions in the $z$-representation can be obtained by constructing the generating functions

\beqn 
G^+(t,z) & = & \sum_{n} \phi^+_n(z)\frac{t^n}{\sqrt {n!}}
=\langle z | \re^{\hud t} | \phi^+_0 \rangle, \nonumber \\
G^-(t,z) & = & \sum_{n} \phi^-_n(z) \frac{t^n}{\sqrt n!}
=\langle z | \re^{\hvd t} | \phi^-_0 \rangle,
\eeqn
that is 
\beqn 
G^+(t,z) & = & 
\pi^{-1/4}
\re^{-\frac {z^2}2 + 2 z \frac {t}{\sqrt 2} -\left ( \frac {t}{\sqrt 2} \right)^2}
\nonumber \\
G^-(t,z)& = & G^+(t,z^*). 
\eeqn 
Using Eq.(\ref{hermite}) we obtain

\beqn
\phi_n^+(z) & = &\frac {1}{\sqrt{n! 2^n \sqrt{\pi}} }H_n(z), \nonumber \\
\phi_n^-(z) & = & (\phi_n^+(z))^*.
\eeqn 

In this representation, the relations of completeness and orthogonality are given by:

\beqn
\sum_{n=0}^\infty ((\phi_n^-(z))^* \phi_n^+(z')+(\phi_n^+(z))^* \phi_n^-(z'))& = &\delta(z-z'), \nonumber \\
\int_{-\infty}^\infty (\phi_n^\mp(z))^* \phi_m^\pm(z) {\rm {dz}}&=&\delta_{nm}.
\eeqn 

In the representation of coordinates 

\begin{equation}
|z \rangle = (2 \pi)^{1/4} \sum_n \left( \phi_n^-(z) |\phi_n^+ \rangle +\phi_n^+(z) |\phi_n^- \rangle \right),
\end{equation}
with 

\beqn 
\mathds 1 =\int_{-\infty}^\infty | z \rangle \langle z | {\rm {dz}}, ~~
\langle z | z' \rangle = \delta(z-z'). 
\eeqn 

The momentum representation can be constructed in the usual form by performing the Fourier Transform of $\phi_n^\pm(z)$:

\beqn 
\mathcal F [\phi_n^\pm](p_z)= \inte \re^{\uni p_z z} \phi_n^\pm(z) {\rm dz }= \uni^n \phi_n^\pm(p_z) ,
\eeqn 
so that 

\begin{equation}
|p \rangle = (2 \pi)^{1/4} \sum_n \left( \phi_n^-(p) |\phi_n^+ \rangle +\phi_n^+(p) |\phi_n^- \rangle \right),
\end{equation}
and 
\beqn 
\mathds 1 =\int_{-\infty}^\infty | p_z \rangle \langle p_z | {\rm {d p_z}}, ~~
\langle p_z | p_z' \rangle = \delta(p_z-p_z'). 
\eeqn

To deal with the continuous spectrum, the hamiltonian of Eq.(\ref{hpos}) can be written as 

\beqn
u \frac{\rd \phi(u)}{\rd u}= \nu \phi(u),
\label{hposc}
\eeqn
with $\nu=-\uni \frac{E}{\hbar |\Om|}- \frac 12 $.
The solutions of Eq. (\ref{hposc}) on $\Phi^\times$ are the generalized functions 

\beqn
\phi_\pm(u) & = & u_\pm^\nu,
\eeqn 
with 

\beqn
s_+^\nu & = &\left \{ 
\begin{array}{ll}
s^\nu  & s \ge 0\\
0 & s < 0
\end{array}
\right .
\nonumber \\
s_-^\nu &=& \left \{ 
\begin{array}{ll}
0 & s \ge 0\\
|s|^ \nu & s<0
\end{array}
\right .
\eeqn

To obtain the coordinate representation we shall use the framework of the Bilateral Mellin Transformation \cite{wolf}, which is a generalization of the expansion in Series of Taylor.

The generating function, $G(t,z/\sqrt{2})$, can be written as 

\beqn 
G\left (t,z \right )=\frac 1{\sqrt{2 \pi}} \sum_s \int_{\mathbb R} \phi_\pm^\lambda \left ( z \right ) t_\pm ^{\uni \lambda-\frac 12 } {\rm d \lambda}.
\eeqn 
By inverting the previous equation and using that \cite{wolf}

\beqn 
\frac 1{2 \pi} \sum_s \int_{\mathbb R} t_s ^{\uni \lambda-\frac 12 } t_{s'} ^{-\uni \lambda'-\frac 12 } {\rm d t}= \delta(\lambda-\lambda') \delta_{s s'},
\eeqn 
we obtain

\beqn
\phi_+^\lambda(z) &=& \mathcal C \int_0^{\infty} {\rm dt} ~ t^{-\uni \lambda-\frac 12 } G\left (t,z \right )
\nonumber \\
&=& \mathcal C ~\Gamma(\nu+1) D_{-\nu-1}(- z \sqrt{2})
\nonumber \\
\phi_-^\lambda(z)
&=& \mathcal C \int_0^{\infty} {\rm dt} ~ t^{-\uni \lambda-\frac 12 } G\left (-t,z \right ) \nonumber \\
&=& \mathcal C ~ \Gamma(\nu+1) D_{-\nu-1}(~\sqrt{2 }z).
\eeqn 
Moreover \cite{chu2} 

\beqn
\int_{-\infty}^{\infty} (\phi_\pm^E(z))^* \phi_\pm^{E'}(z) {\rm dz}&=&
\delta(E- E') \nonumber \\
\sum_{s=\pm } \int_{-\infty}^{\infty}(\phi_s^E(z))^* \phi_s^{E}(z') {\rm dE}&=&
\delta(z-z').
\label{ort-comp}
\eeqn

There is also a set of solutions corresponding to the hamiltonian of Eq.(\ref{hpos}), that is

\beqn
v \frac{\rd \phi(v)}{\rd v}= \lambda \phi(v),
\label{hnegc}
\eeqn
with $\lambda=\uni \frac{E}{\hbar |\Om|}- \frac 12= \nu^*=-(\nu+1)$. That is,  we change \\ $\nu \rightarrow -(\nu+1)$ and $E \rightarrow -E$, so that $h |\phi \rangle=E|\phi \rangle \rightarrow h |\eta \rangle = -E |\eta\rangle$. Thus the solutions of Eq.(\ref{hnegc}) can be given in terms of the solutions of Eq.(\ref{hposc}) as

\beqn 
\eta^E_\pm(z)= (\phi^E_\pm(z))^*.
\eeqn
We can express Eq.(\ref{ort-comp}) as 

\beqn
\int_{-\infty}^{\infty} \eta_\pm^E(z)) \phi_\pm^{E'}(z) {\rm dz}&=&
\delta(E- E') \nonumber \\
\int_{-\infty}^{\infty} \left( \eta_+^E(z) \phi_+^{E}(z') + \phi_-^E(z) \eta_-^{E}(z') \right) {\rm dE}&=&
\delta(z-z').\nonumber \\
\label{ort-comp}
\eeqn

In the coordinate representation, we have 
$\phi_\pm^E(z)= \langle z |\phi_\pm^E\rangle$ and $\eta_\pm^E(z)= \langle z | \eta_\pm^E\rangle=$. Consequently: 

\beqn 
|z \rangle = \frac 1{2 \sqrt {\pi}} \sum_{s=\pm }
\left( \eta_s^E(z) |\phi_s^E \rangle+ \phi_s^E(z) |\eta_s^E \rangle \right),
\eeqn 
so that $\langle z| z' \rangle=\delta(z-z')$.

\subsection*{Boundaries I-II and III-IV.}

In both boundaries, I-II and III-IV, $\Om$ takes the value $\Om=0$.
When $\Omega=0$ and $\omega -(\alpha+\beta) \ne 0$, the problem reduces to that of a free particle of energy $E$, that is

\beqn
-\frac{\hbar^2}{2 m  }  \frac {{\rm d}^2 \phi(x)}{{\rm d} x^2} = E ~\phi(x),
\label{hosc1}
\eeqn
so that the generalized eigenfunction can be written as 
$\phi(x)=A {\rm e}^{i k x}+B {\rm e}^{-i k x}$, with 
$k= \sqrt{\frac{2 E}{\hbar (\omega-\alpha-\beta) b_0^2}}$.

For the case $E=0$, we have another solution, $\phi(x)=c_0+c_1 x$.


\section*{Appendix B}

The eigenfunctions and the spectrum in the boundary between regions I-III can be obtained in a limit procedure from region I and region III \cite{mannheim1,mannheim2}.
To see this, consider the hamiltonian of Eq. (\ref{hxp}) replacing $\omega - \alpha - \beta$ by $\varepsilon$:

\beqn
H^\times(\varepsilon,\alpha,\beta) & = &
\frac 1 2 \hbar (\varepsilon +2( \alpha +\beta) ) \left( \frac {\hat x} {b_0}\right)^2 \nonumber \\
& &+ \frac 1 2 \hbar \varepsilon \left( \frac {b_0~ \hat{p}} \hbar \right)^2
\nonumber \\
& &+  \hbar \frac{(\alpha-\beta)}{2} \left( 2~\hat{x} \frac{ {\bf i}}{\hbar} \hat{p}+1 \right).
\label{varepsilon}
\eeqn
Its generalized eigenfunctions are given by 

\beqn 
\wid \phi_{n, \varepsilon}(x)= 
\re^{ \tfrac{x^2}{2 b_0^2 } \frac{(\al-\be)}{\varepsilon} } \re^{ -\tfrac{x^2}{2 b_0^2 } 
\frac{ r(\varepsilon)} {\varepsilon}}~
H_n \left (\frac{x}{b_0} \sqrt{\frac{r(\varepsilon)}{\varepsilon }} \right),
\label{phinl}
\eeqn 
with $r(\varepsilon)=\sqrt{(\alpha-\be)^2+2 \varepsilon  (\al+\be)+\varepsilon^2}$.
As shown before, the
solutions for $H^\times(\varepsilon,\alpha,\beta) \wid \phi_{n, \varepsilon}(x)= E_{n, \varepsilon} \wid \phi_{n, \varepsilon}(x) $,
with
$E_{n, \varepsilon}= \left (n+\frac{1}{2} \right) r(\varepsilon)$.


Let us further introduce a new parameter $G$ defined by $G=\frac {r(\varepsilon)} {\varepsilon}$.
The limit $\varepsilon \rightarrow 0$ should be replaced by the limit $G\rightarrow \infty$. 
In terms of $G$:

\beqn 
\varepsilon_\pm = \frac{(\alpha +\beta)}{G^2-1} \pm
\frac{\sqrt{4 \alpha  \beta +G^2 (\alpha -\beta )^2}}{G^2-1}.
\eeqn 

For $\varepsilon=\varepsilon_+$, we have  
 
\beqn
\wid \phi_{n, G}^{(1)}(x)= 
C \tau^{-1} 
\re^{-\frac{x^2 G}{2 b_0^2} 
\left( \sqrt{1+ \frac{4 \al \be}{ G^2(\al-\be)^2} }-1 \right)}
H_n\left(\frac x {b_0} \sqrt{G} \right),\nonumber\\
\eeqn 
where $C=\frac{2^{-n}G^{-n/2}}{\sqrt{n!}}$.
Notice that, when $G \rightarrow \infty$, we obtain 
\beqn
\wid \phi_{n, G}^{(1)}(x)
\rightarrow \tau^{-1 }  \frac{x^{n}}{\sqrt{n!}}.
\eeqn 
Now, we shall take $\varepsilon = \varepsilon_-$. Then $\wid \phi_{n, \varepsilon}(x)$ can be expressed in terms of $G$ as
\beqn 
\wid \phi_{n, G}^{(2)}(x)= C \tau^{-1} 
\re^{-\frac{x^2 G}{2 b_0^2} 
\left( \sqrt{1+ \frac{4 \al \be}{ G^2(\al-\be)^2} }+1 \right)}
H_n\left(\sqrt{G} x\right),\nonumber\\
\eeqn
we shall call
\beqn 
\phi_{n, G}^{(2)}(x)= \tau \wid \phi_{n, G}^{(2)}(x).
\eeqn 
It can be proved that

\beqn 
\mathcal{F}[\phi_{n, G}^{(1)}](w)=q_{n}(G)\phi_{n, G}^{(2)}
\left (\tfrac 12 \tfrac{\uni (\alpha -\beta )}{\sqrt{\alpha \beta }}w \right),
\eeqn 
where 

\beqn 
q_{n}(G)=
\frac{(2 \sqrt{\al \be})^n}{(\al-\be)^n} \frac{1}{G^{n +\frac 12} 
\left( 1-\sqrt{1+ \frac{4 \al \be}{ G^2(\al-\be)^2} }\right)^{n +\frac 12 }},
\nonumber \\
\eeqn 
Taking $a= \uni \frac{\alpha - \beta}{\sqrt{\alpha \beta}}$ and using the property of the Fourier Transform $\mathcal{F}(f(a t))(w)= \tfrac{1}{|a|}\mathcal{F}(f(t))(\frac{w}{a})$, we obtain
\begin{eqnarray*}
q_{n}(G)\phi_{n, G}^{(2)}(w)&=& \mathcal{F}[\phi_{n, G}^{(1)}]\left (w/a \right)\\
&=& |a| (\tfrac{1}{|a|} \mathcal{F}[\phi_{n, G}^{(1)}(x)](w/a))\\
q_{n}(G)\phi_{n, G}^{(2)}(w) &\rightarrow & |a| \left ( \tfrac{1}{|a|} \mathcal{F}[\tfrac{x^n}{\sqrt{n!}}](\tfrac{w}{a}) \right)\\
&=& c_n \frac{(-1)^{n}\delta^{(n)}(w)}{\sqrt{n!}}
\end{eqnarray*}
Then
\beqn
\wid \phi_{n, G}^{(2)}(x)\rightarrow \widetilde{\phi}_n^-(x).
\eeqn
The same study can be made for $H^\times_c$, so that 

\beqn
\wid \phi_{n, G}^{(1)}(x) & \rightarrow & \ovl{\psi}_n^-(x), \nonumber \\
\wid \phi_{n, G}^{(2)}(x) & \rightarrow & \ovl{\psi}_n^+(x).
\eeqn

The eigenfunctions of Region I, with discrete positive spectrum, have as punctual limit the eigenfunctions of the boundary I-III with positive eigenvalues.  
In the same way, the eigenfunctions of Region III, with discrete negative spectrum, have as punctual limit the eigenfunctions of the boundary I-III with negative eigenvalues.   

\section*{Appendix C}

To study the coalescence of the discrete eigenvalues and eigenfunctions between Regions I-II and III-IV, we have to look at the Hamiltonian of Eq.(\ref{hxp}). If $\omega^2-4\alpha\beta=0$, we have

\beqn 
H^\times & = & 
\frac 12 \hbar \frac{ (w+2 \beta)^2}{8 \beta } \hat x^2 
\nonumber \\ & & 
+\frac 12 \hbar  \frac{(w-2 \beta )^2 }{8 \beta } \left( \frac{\hat p }{\hbar} \right)^2
\nonumber \\ & & 
+\hbar \frac{(w^2-4 \beta ^2)}{8 \beta } 
\left (2 \hat x \frac{\uni}{\hbar} \hat p + 1 \right ),
\eeqn 
and

\beqn 
H^\times_c & = & 
\frac 12 \hbar \frac{ (w+2 \beta)^2}{8 \beta } \hat x^2 
\nonumber \\ & & 
+\frac 12 \hbar  \frac{(w-2 \beta )^2 }{8 \beta } \left( \frac{\hat p }{\hbar} \right)^2
\nonumber \\ & & 
-\hbar \frac{(w^2-4 \beta ^2)}{8 \beta } 
\left (2 \hat x \frac{\uni}{\hbar} \hat p+1 \right ),
\eeqn 

The discrete eigenvalues, $E_n \rightarrow 0$ when $\om^2-4 \al \be \rightarrow 0$. We shall look for the eigenfunctions  of $H^\times$ and $H^\times_c$ for $E=0$:

\beqn
H^\times \wid \phi(x)   & = & 0, \nonumber \\
H_c^\times \ovl \psi(x) & = & 0.
\eeqn 
The solutions to the problem are given by
\beqn
\widetilde{\phi}(x) & = & \left(c_1 x+c_0\right) e^{-\frac{x^2 (w+2 \beta)}{2 (w-2 \beta )}} \nonumber \\
\overline{\psi}(x)  & = & \left(d_1 x+d_0\right) e^{~\frac{x^2 (w+2 \beta)}{2 (w-2 \beta )}}.
\eeqn

The behaviour of the eigenfunctions at the border between I-II and III-IV can be obtained by using a limit procedure \cite{mannheim1,mannheim2} from the discrete solutions in each region. 

In Region I, we shall take $\om^2- 4 \al \be = \epsilon^2$ and  we shall solve the equation

\beqn 
H^\times \psi_{n, \varepsilon}(x)= E_{n, \varepsilon}\psi_{n, \varepsilon}(x).
\eeqn 
We obtain
\beqn
\wid \phi_{n, \varepsilon}(x) & = & 
\mathcal{N}_{n}(\varepsilon)
\re^{-\frac{4 | \beta |^2 -\om^2(1-\epsilon^2)}{(\om +2 |\beta|)^2 -\epsilon^2 \om^2} 
\frac{x^2}{2 b_0^2} } 
H_n\left( x \sg(\varepsilon) \right), 
\nonumber \\
E_{n, \varepsilon} & = & \hbar \varepsilon (n+\tfrac{1}{2}),
\eeqn 
with $\sg(\varepsilon)=\sqrt{ \tfrac{ 4 |\be| \epsilon }
{(w+ 2 |\be |)^2-4\epsilon^2 \om^2} }$ and 

\beqn
\mathcal{N}_{n}^{2}(\varepsilon) \sim  
\tfrac
{\left(\left(2 \beta +\epsilon \right)^2-w^2\right)^{n}}
{\frac{1}{2} \left(\beta  \left(-\left((-1)^n-1\right)\right) \epsilon +(-1)^n+1\right)P_{g(n)}(\varepsilon)},
\eeqn
where $P_{g(n)}(\varepsilon)= \sum_{k=0}^{g(n)}t_{k}\varepsilon^{k}$, $g(n)=\tfrac{1}{2}(-1 + (-1)^n + 2 n)$ and $t_{0}\sim (w^2-4 \beta ^2)^{g(n)}$. Notice that
$$\mathcal{N}_{2n}(\varepsilon)\sim \varepsilon^{0}, ~~~
\mathcal{N}_{2n+1}(\varepsilon) \sim \varepsilon^{-1/2}, $$
then

\beqn 
\wid \phi_{2n, \varepsilon  }(x)  & \rightarrow &   \re^{-\frac{x^2 (w+2 \beta)}{2 (w-2 \beta )}},\nonumber \\
\wid \phi_{2n+1, \varepsilon}(x)  & \rightarrow & x \re^{-\frac{x^2 (w+2 \beta)}{2 (w-2 \beta )}},
\eeqn 
so that 
\beqn 
\wid \phi_{\varepsilon}(x) & \rightarrow & (c_0+ c_1 x)~ \re^{-\frac{x^2 (w+2 \beta )}{2 (w-2 \beta )}} \nonumber \\
\ovl \psi_{\varepsilon}(x) & \rightarrow & (d_0+ d_1 x)~ \re^{~ \frac{x^2 (w+2 \beta)}{2 (w-2 \beta )}} \nonumber \\
E_{n, \varepsilon}  & \rightarrow & 0.
\label{reg1}
\eeqn 

Let us procedure in the same form in Region II. As $\Om^2 <0$, we shall take  
$\omega^2 -  4 \alpha \beta = -\epsilon^2$, so that the discrete spectrum its given by $E_{n,\epsilon}^\pm=\pm \uni \hbar \sqrt{\varepsilon} \left( n +\frac 12 \right)$. The eigenfunctions, in terms of $\varepsilon$, are given by
\beqn
\wid \phi_{n, \varepsilon}^\pm(x) & = & 
\mathcal{N}_{n}^\pm (\varepsilon)
\re^{-\frac{4 | \beta |^2 -\om^2(1-\epsilon^2)}{(\om +2 |\beta|)^2 -\epsilon^2 \om^2} 
\frac{x^2}{2 b_0^2} } 
H_n\left( \re^{\pm \uni \pi/4} x \sg(\varepsilon) \right), 
\nonumber \\
\wid E_{n, \varepsilon}^\pm & = & \pm \uni \hbar \varepsilon (n+\tfrac{1}{2}), \nonumber \\
\ovl \psi_{n, \varepsilon}^\mp(x) & = & 
\mathcal{N}_{n}^\pm (\varepsilon)
\re^{~\frac{4 | \beta |^2 -\om^2(1-\epsilon^2)}{(\om +2 |\beta|)^2 -\epsilon^2 \om^2} 
\frac{x^2}{2 b_0^2} } 
H_n\left( \re^{\mp \uni \pi/4} x \sg(\varepsilon) \right), 
\nonumber \\
\ovl E_{n, \varepsilon}^\pm & = & \mp \uni \hbar \varepsilon (n+\tfrac{1}{2}).
\eeqn 

In this case, normalization constants are given by
$$(\mathcal{N}_{n}^{\pm})^{2}(\epsilon)\sim  \tfrac{\left(\left(2 \beta +\sqrt{\epsilon }\right)^2-w^2\right)^{n}}{\frac{1}{2} \left(\beta  \left(-\left((-1)^n-1\right)\right) \sqrt{\epsilon }+(-1)^n+1\right)P_{g(n)}(\varepsilon)}$$
where $P_{g(n)}(\epsilon)$ is a polynomial with order  $g(n)=\tfrac{1}{2}(-1 + (-1)^n + 2 n)$ in $\epsilon$ and independent term $t_{0}\sim (w^2-4 \beta ^2)^{g(n)}$. As before
$$(\mathcal{N}_{n}^{\pm})^{2}(\epsilon)\sim \epsilon^{0}, ~~~
(\mathcal{N}_{n}^{\pm})^{2}(\epsilon)\sim \epsilon^{-1/4},$$
and 

\beqn 
\wid{\phi}_{2n,\epsilon}^\pm(x)   & \rightarrow &   e^{-\frac{x^2 (w+2 \beta)}{2 (w-2 \beta )}}, \nonumber \\
\wid{\phi}_{2n+1,\epsilon}^\pm(x) & \rightarrow & x e^{-\frac{x^2 (w+2 \beta)}{2 (w-2 \beta )}},
\eeqn
so that finally:

\beqn 
\wid \phi_{\varepsilon}(x)^\pm & \rightarrow & (c_0+ c_1 x)~ \re^{-\frac{x^2 (w+2 \beta )}{2 (w-2 \beta )}} \nonumber \\
\ovl \psi_{\varepsilon}(x)^\pm & \rightarrow & (d_0+ d_1 x)~ \re^{~ \frac{x^2 (w+2 \beta)}{2 (w-2 \beta )}} \nonumber \\
E_{n, \varepsilon}  & \rightarrow & 0.
\label{reg2}
\eeqn 

Clearly, from Eqs.(\ref{reg1}) and (\ref{reg2}), it can be concluded that the boundary I-II includes EPs.

The procedure we have applied to the limit of Regions I and II can be implemented in the limit between regions III and IV. It is straightforward to prove that the boundary III-IV, also includes EPs.
\section*{Acknowledgments}

This work was partially supported by the National Research Council of Argentine (CONICET) (PIP 0616) and by the Agencia Nacional de Promoci\'on Cient\'\i fica (ANPCYT)  of Argentina.

\end{document}